\pdfoutput=1

\documentclass[preprint,showkeys,secnumarabic,amsfonts,showpacs,amsmath,amssymb]{revtex4}
\usepackage[dvips]{color}
\usepackage{epsfig}
\usepackage{amsmath}
\usepackage{graphicx}

\begin{document}

\title{\bf Observational constraints in scalar tensor theory with tachyonic potential}
\author{Hossein Farajollahi}
\email{hosseinf@guilan.ac.ir}
\affiliation{$^1$Department of Physics, University of Guilan, Rasht, Iran}
\affiliation{$^2$ School of Physics, University of New South Wales, Sydney, NSW, 2052, Australia}

\author{Amin Salehi}
\email{a.salehi@guilan.ac.ir}
\affiliation{Department of Physics, University of Guilan, Rasht, Iran}

\author{Asieh Shahabi}
\email{ashahabi@guilan.ac.ir}
\affiliation{Department of Physics, University of Guilan, Rasht, Iran}

\date{\today}

\begin{abstract}
We study the dynamics of the scalar tensor cosmological model in the presence of tachyon field. In an alternative approach, in two exponential and power law form of the scalar field functions in the model, field equations are solved by simultaneously best fitting the model parameters with the most recent observational data. This approach gives us an
observationally verified interpretation of the dynamics of the universe. We then discuss the best fitted of equation of state parameter, the statefinder parameters and the reconstructed scalar field in the model.

\end{abstract}



\maketitle

\section{Introduction}

Recently, the observations of high redshift type Ia
supernovae, the surveys of clusters of galaxies \cite{Reiss}--\cite{Riess2}, Sloan digital sky survey ({\bf
SDSS})~\cite{Abazajian} and Chandra X--ray observatory~\cite{Allen} reveal the universe accelerating expansion
and that the density of matter is very much less than the critical density. Also the
observations of Cosmic Microwave Background (CMB)
anisotropies \cite{Bennett} indicate that the universe is flat and the total energy
density is very close to the critical one \cite{Spergel}. The observations though determines basic cosmological parameters
with high precisions and strongly indicates that the universe
presently is dominated by a smoothly distributed and slowly
varying dark energy (DE) component, but at the same time they poses a serious
problem about the origin of DE \cite{Tsujikawa}. A dynamical equation of state ( EoS) parameter that is connected directly to the evolution of the energy density in the universe and indirectly to the expansion of the Universe can be regarded as a suitable parameter to explain the acceleration and the origin of DE \cite{Seljak}--\cite{Setare}. In scalar-tensor theories \cite{Sahoo}--\cite{Nojiri3}, interaction of the scalar field with matter ( for example in chameleon cosmology) \cite{Setare1}--\cite{Dimopoulos} or with geometry in Brans-Dicke (BD) cosmological models \cite{Damouri}--\cite{Biswass} can be used to interpret the late time acceleration.

On the other hand, by using the well-known geometric variables, Hubble parameter and deceleration parameter together with the new geometrical variables, the cosmological
diagnostic pair $\{r, s\}$ ( or statefinder parameters)\cite{Sahni}, the acceleration expansion of the universe and differentiation among the cosmological models can be explained in order to better fit the observational data. The importance of the statefinder parameters to distinct DE cosmological models is best realized, in particular, when considering the increased accuracy of the observational data during the last few years and
generality of the DE models. These parameters, in a natural next step beyond the well known geometric variables, are to differentiate the expansion dynamics
with higher derivatives of the scale factor and to explore a series of DE cosmological models,
including $\Lambda$ cold dark matter ($LCDM$), quintessence, coupled quintessence, Chaplygin gas, holographic dark energy
models, braneworld models, and so on \cite{Alam}--\cite{faraj2}. Moreover, since the cosmic acceleration affects the expansion history of the
universe, to understand the true nature of the driving force,
mapping of the cosmic expansion of the universe is very
crucial~\cite{Linder}. Hence, one requires various observational
probes in different redshift ranges to understand the expansion
history of the universe. One of these tests is the difference in distance modulus measurement of
 type Ia supernovae that helps us to testify the cosmological models.

In this paper, we investigate the dynamics of the scalar-tensor theory with tachyon potential where an scalar field function $F(\phi)$ coupled to the curvature and matter lagrangian. The model allows scalar field function that might be light on cosmological
scales, to couple to matter much more strongly than gravity does, and still satisfies the current
experimental and observational constraints. The cosmological value
of such a field evolves over Hubble time-scales and could
potentially cause the late--time acceleration of our universe.

 The approach we used in this paper is simultaneously solving the field equations and best fitting the model parameters and initial conditions with the most recent observational date for distance modulus using chi-squared method. Sec. two is devoted to a detailed formulation of the cosmological model. In sec. three we solve the system of field equations by simultaneously best fitting the model parameters and initial conditions with the observational data for distance modulus. In Sec. four, we examine the behavior of the best fitted effective EoS parameter of the model and also perform a statefinder diagnostic for the model and analyze the evolving trajectories of the model in the statefinder parameter plane. In Sec. five, we present summary and remarks.

\section{The model}

The model is presented by the action,
\begin{eqnarray}\label{ac1}
S=\int[F(\varphi)R-V(\varphi)\sqrt{1-\varphi_{,\mu}\varphi^{,\mu}}+F(\varphi)L_{m}]\sqrt{-g}d^{4}x
\end{eqnarray}
where $R$ is the Ricci scalar, $G$ is the Newtonian constant gravity, and the second term in the action is the
tachyon potential. The $F(\varphi)$ and $V(\varphi)$ are analytic function of the scalar field. The last term in the Lagrangian brings
about the nonminimal interaction between the matter and the scalar field. The variation of action (\ref{ac1}) with
respect to the metric tensor components in a spatially flat Friedmann-Robertson-Walker cosmology
yields the field equations:
\begin{eqnarray}
&&F[2\dot{H}+3H^2]+
2[\dot{\varphi}\frac{dF}{d\varphi}]H
+[\dot{\varphi}^{2}\frac{d^{2}F}{d\varphi}^{2}+\ddot{\varphi}\frac{dF}{d\varphi}]-\frac{1}{2}V\sqrt{1-\dot{\varphi^{2}}}+\gamma\rho_{m}
=0,\label{ac9}\\
&&6FH^{2}+6[\dot{\varphi}\frac{dF}{d\varphi}]H
+\frac{1}{2}(\dot{\varphi}^{2})-\frac{V}{\sqrt{1-\dot{\varphi^{2}}}}-\rho_{m}=0 \label{ac10}
\end{eqnarray}
where $\epsilon=1-3\gamma$ and $\gamma$ is the EoS parameter for the matter filled the universe. Variation with respect to the scalar field $\phi$ gives
\begin{eqnarray}\label{ac11}
\ddot{\varphi}+(3H\dot{\varphi}+\frac{V^{'}}{V})(1-\dot{\varphi^{2}})=+\frac{\epsilon\rho_{m}F^{'}}{V}(1-\dot{\varphi^{2}})^{\frac{3}{2}}+\frac{F^{'}(6\dot{H}+12H^{2})}{V}(1-\dot{\varphi^{2}})^{\frac{3}{2}}
\end{eqnarray}
where prime is derivative with respect to the scalar field. From the above field equations, conservation equation is obtained as
\begin{eqnarray}
\dot{\rho_{m}}+3H(1+\gamma)\rho_{m}=-\rho_{m}\frac{\dot{F}}{F}(1+\epsilon)
\end{eqnarray}
which readily integrates to yield,
 \begin{eqnarray}
\rho_{m}=\frac{M}{a^{3(1+\gamma)}F^{(1+\epsilon)}}\label{conserv}
\end{eqnarray}
with $M$ is a integration constant and eq (\ref{conserv}) is the energy constraint corresponding to the (0,0)-Einstein equation. In the following we assume that the matter presented in the universe is cold dark matter, $\gamma=0$. We now study the structure of the dynamical system by introducing the following dimensionless variables,
\begin{eqnarray}\label{ac13}
X=\frac{V}{FH^{2}},\ \
Y=\dot{\varphi},\ \ Z=H,\ \ U=\frac{\rho_{m}}{6H^{2}}
\end{eqnarray}

In the following we consider two cases:

{\bf Exponential functions: $F(\phi)\propto e^{\alpha \phi}$ and $V(\phi)\propto e^{\beta \phi}$}

Using equations (\ref{ac9})-(\ref{ac11}), the evolution equations of these variables become,
\begin{eqnarray}
X^{'}&=&\frac{(\beta-\alpha)XY}{Z}-2X\frac{\dot{H}}{H^{2}}\label{ac14}\\
Y^{'}&=&-3Y(1-Y^{2})-\frac{\beta (1-Y^{2})}{Z}+\frac{\alpha }{X Z}(6\frac{\dot{H}}{H^{2}}+12)(1-Y^{2})^{\frac{3}{2}}+6\epsilon\alpha(1-Y^{2})^{\frac{3}{2}}\frac{U}{X Z}\label{ac15}\\
Z^{'}&=&Z\frac{\dot{H}}{H^{2}}\\
U^{'}&=&-3U-\frac{2\alpha Y U}{Z}-2U\frac{\dot{H}}{H^{2}}\label{u14}
\end{eqnarray}
where prime in here and from now on is taken to be derivative with respect to $N = ln (a)$ and
\begin{eqnarray}\label{hdotexp}
\frac{\dot{H}}{H^{2}}&=&\frac{1}{(XZ^{2}+{3\alpha^{2}(1-Y^{2})^{\frac{3}{2}}})}\big[-\frac{3}{2}XZ^{2}+\frac{\alpha YZX}{2}(1-3Y^{2} ) +\frac{\beta\alpha X(1-Y^{2})}{2}
\\&-&\nonumber3\alpha^{2}U(1-Y^{2})^{\frac{3}{2}}-{6\alpha^{2}(1-Y^{2})^{\frac{3}{2}}}-\frac{\alpha^{2}Y^{2}X}{2}+
\frac{X^{2}Z^{2}(1-Y^{2})^{\frac{1}{2}}}{4}]
\end{eqnarray}
We also have the constraint equation in terms of the new variables as
\begin{eqnarray}\label{ac17}
1+\frac{\alpha Y}{Z}-\frac{X}{6\sqrt{1-Y^{2}}}-U=0
\end{eqnarray}
By imposing the constraint equation, (\ref{ac17}), equations (\ref{ac14})-(\ref{u14}) reduce to
\begin{eqnarray}
X^{'}&=&(\beta-\alpha)XYZ-2X\frac{\dot{H}}{H^{2}}\label{ac19}\\
Y^{'}&=&-3Y(1-Y^{2})-\frac{\beta (1-Y^{2})}{Z}+\frac{\alpha }{X Z}(6\frac{\dot{H}}{H^{2}}+12)(1-Y^{2})^{\frac{3}{2}}
\\&+&\nonumber 6\alpha(1-Y^{2})^{\frac{3}{2}}\frac{(1+\frac{\alpha Y}{Z}-\frac{X}{6\sqrt{1-Y^{2}}})}{X Z}\label{ac20}\\
Z^{'}&=&Z\frac{\dot{H}}{H^{2}}
\end{eqnarray}

{\bf Power law functions: $F(\phi)\propto \phi^\alpha$ and $V(\phi)\propto \phi^\beta$}

By taking the scalar function $F(\phi)$ and potential $V(\phi)$ in power law form the dynamical equations become

\begin{eqnarray}
X^{'}&=&\frac{(\beta-\alpha)XY}{Z(XZ^{2})^{\frac{1}{\beta-\alpha}}}-2X\frac{\dot{H}}{H^{2}}\label{ac114}\\
Y^{'}&=&-3Y(1-Y^{2})-\frac{\beta (1-Y^{2})}{Z(XZ^{2})^{\frac{1}{\beta-\alpha}}}+\frac{\alpha  }{X Z(XZ^{2})^{\frac{1}{\beta-\alpha}}}(6\frac{\dot{H}}{H^{2}}+12)(1-Y^{2})^{\frac{3}{2}}
\\&+&\nonumber6\alpha (1-Y^{2})^{\frac{3}{2}}\frac{U}{X Z(XZ^{2})^{\frac{1}{\beta-\alpha}}}\label{ac15}\\
Z^{'}&=&Z\frac{\dot{H}}{H^{2}}\\
U^{'}&=&-3U-\frac{2\alpha Y U}{Z(XZ^{2})^{\frac{1}{\beta-\alpha}}}-2U\frac{\dot{H}}{H^{2}}\label{upower}
\end{eqnarray}
where
\begin{eqnarray}\label{hdotpower}
\frac{\dot{H}}{H^{2}}&=&\frac{1}{[XZ^{2}(X^{2}Z^{\frac{4}{\beta-\alpha}})+{3\alpha^{2}(1-Y^{2})^{\frac{3}{2}}}]}
\big[-\frac{3}{2}XZ^{2}(X^{2}Z^{\frac{4}{\beta-\alpha}})+\frac{\alpha YZX}{2}(1-3Y^{2} )(XZ^{2})^{\frac{1}{\beta-\alpha}}\nonumber\\
&+&\nonumber\frac{\beta\alpha X(1-Y^{2})}{2}-3\alpha^{2}U(1-Y^{2})^{\frac{3}{2}}-{6\alpha^{2}(1-Y^{2})^{\frac{3}{2}}}\nonumber\\
&-&\nonumber\frac{\alpha (\alpha-1)Y X}{2}+\frac{X^{2}Z^{2}(1-Y^{2})^{\frac{1}{2}}(X^{2}Z^{\frac{4}{\beta-\alpha}})}{4}\big]
\end{eqnarray}
The constraint Friedmann equation now becomes
\begin{eqnarray}\label{ac18}
U=1+\frac{\alpha Y}{Z(XZ^{2})^{\frac{1}{\beta-\alpha}}}-\frac{X}{6\sqrt{1-Y^{2}}}
\end{eqnarray}
Using the constraint (\ref{ac18}), the above equations now reduce to the following equations:
\begin{eqnarray}
X^{'}&=&\frac{(\beta-\alpha)XY}{Z(XZ^{2})^{\frac{1}{\beta-\alpha}}}-2X\frac{\dot{H}}{H^{2}}\label{ac19}\\
Y^{'}&=&-3Y(1-Y^{2})-\frac{\beta (1-Y^{2})}{Z(XZ^{2})^{\frac{1}{\beta-\alpha}}}+\frac{\alpha  }{X Z(XZ^{2})^{\frac{1}{\beta-\alpha}}}(6\frac{\dot{H}}{H^{2}}+12)(1-Y^{2})^{\frac{3}{2}}\\
&+&\nonumber6\alpha (1-Y^{2})^{\frac{3}{2}}\frac{1+\frac{\alpha Y}{Z(XZ^{2})^{\frac{1}{\beta-\alpha}}}-\frac{X}{6\sqrt{1-Y^{2}}}}{X Z(XZ^{2})^{\frac{1}{\beta-\alpha}}}\label{ac20}\\
Z^{'}&=&Z\frac{\dot{H}}{H^{2}}
\end{eqnarray}
In the following we solve the dynamical systems on both cases by best-fitting the model parameters and also initial conditions with the observational data.

\section{Observational best fit by using distance modulus, $\mu(z)$}

The difference between the absolute and
apparent luminosity of a distance object is given by, $\mu(z) = 25 + 5\log_{10}d_L(z)$ where the Luminosity distance quantity, $d_L(z)$ is given by
\begin{equation}\label{dl}
d_{L}(z)=(1+z)\int_0^z{-\frac{dz'}{H(z')}}.
 \end{equation}
With numerical computation, we solve the system of dynamical equations for $X$ and $Y$ and $Z$ in both in
both power law and exponential cases. In addition, to best fit the model parameters and
initial conditions we use the following three auxiliary equations for the luminosity distance
and the hubble parameter
\begin{eqnarray}
 \frac{dH}{dN}&=&H(-\frac{\dot{H}}{H^{2}})\\
\frac{d(d_{L})}{dN}&=&-d_{L}-\frac{e^{-2N}}{H}\\
z=&&-1+e^{-N}.
\end{eqnarray}
From numerical computation one can obtain $H(z)$ which can be used to evaluate $\mu(z)$. To best fit the model for the parameter $\alpha$ and $\beta$, the initial conditions $Y(0)$, $X(0)$, $Z(0)$ with the most recent observational data, the Type Ia supernovae (SNe Ia), we employ the $\chi^2$ method. We constrain the parameters including the initial conditions by minimizing the $\chi^2$ function given as
\begin{equation}\label{chi2}
 \chi^2_{SNe} (\alpha, \beta, X(0), Y(0),Z(0))=\sum_{i=1}^{557}\frac{[\mu_i^{the}(z_i|\alpha, \beta, X(0), Y(0), Z(0)) - \mu_i^{obs}]^2}{\sigma_i^2},
\end{equation}
where the sum is over the SNe Ia sample. In relation (\ref{chi2}), $\mu_i^{the}$ and $\mu_i^{obs}$ are the distance modulus parameters obtained from our model and from observation, respectively, and $\sigma$ is the estimated error of the $\mu_i^{obs}$. From
numerical computation, Table I shows the best best-fitted model parameters in both cases.

\begin{table}[ht]
\caption{Best-fitted model parameters and initial conditions.} 
\centering 
\begin{tabular}{c c c c c c c } 
\hline 
parameters  &  $\alpha$  &  $\beta$ \ & $X(0)$\ & $Y(0)$\ & $Z(0)$ \ & $\chi^2_{min}$\\ [2ex] 
\hline 
Exponential&$-0.23$  & $1.17$ \ & $5.5$\ & $-0.6$\ & $0.91$ \ & $544.530862$ \\
Power law  & $-2.97$  & $-1.82$ \ & $1$\ & $0.96$\ & $0.91$ \ & $546.1139533$ \\
\hline 
\end{tabular}
\label{table:1} 
\end{table}\

Figures 1-4 shows the constraints on the parameters $\alpha$ and $\beta$ at the $68.3\%$, $95.4\%$ and $99.7\%$ confidence levels in both cases of exponential and power law functions.

\begin{tabular*}{2.5 cm}{cc}
\includegraphics[scale=.35]{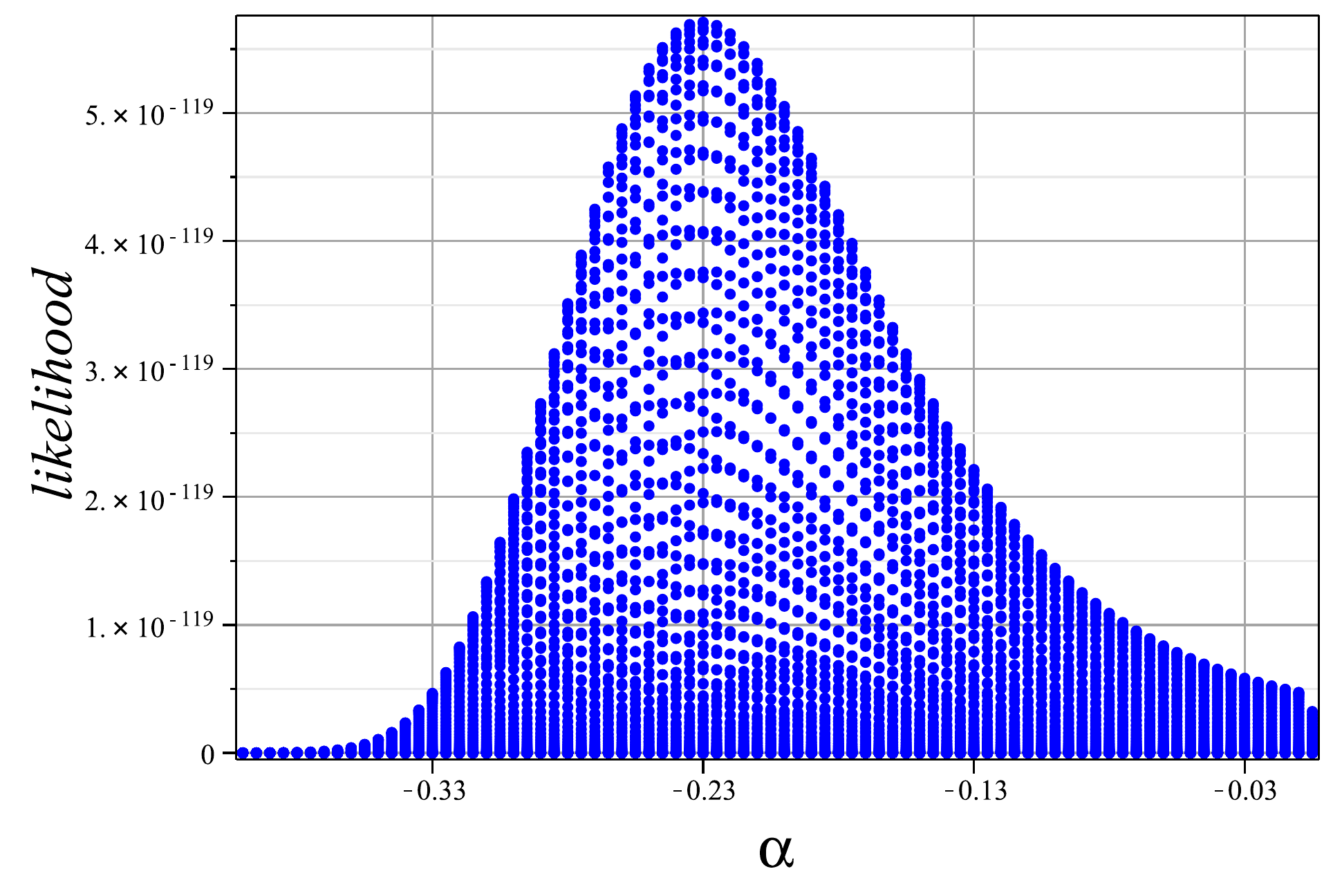}\hspace{0.1 cm}\includegraphics[scale=.35]{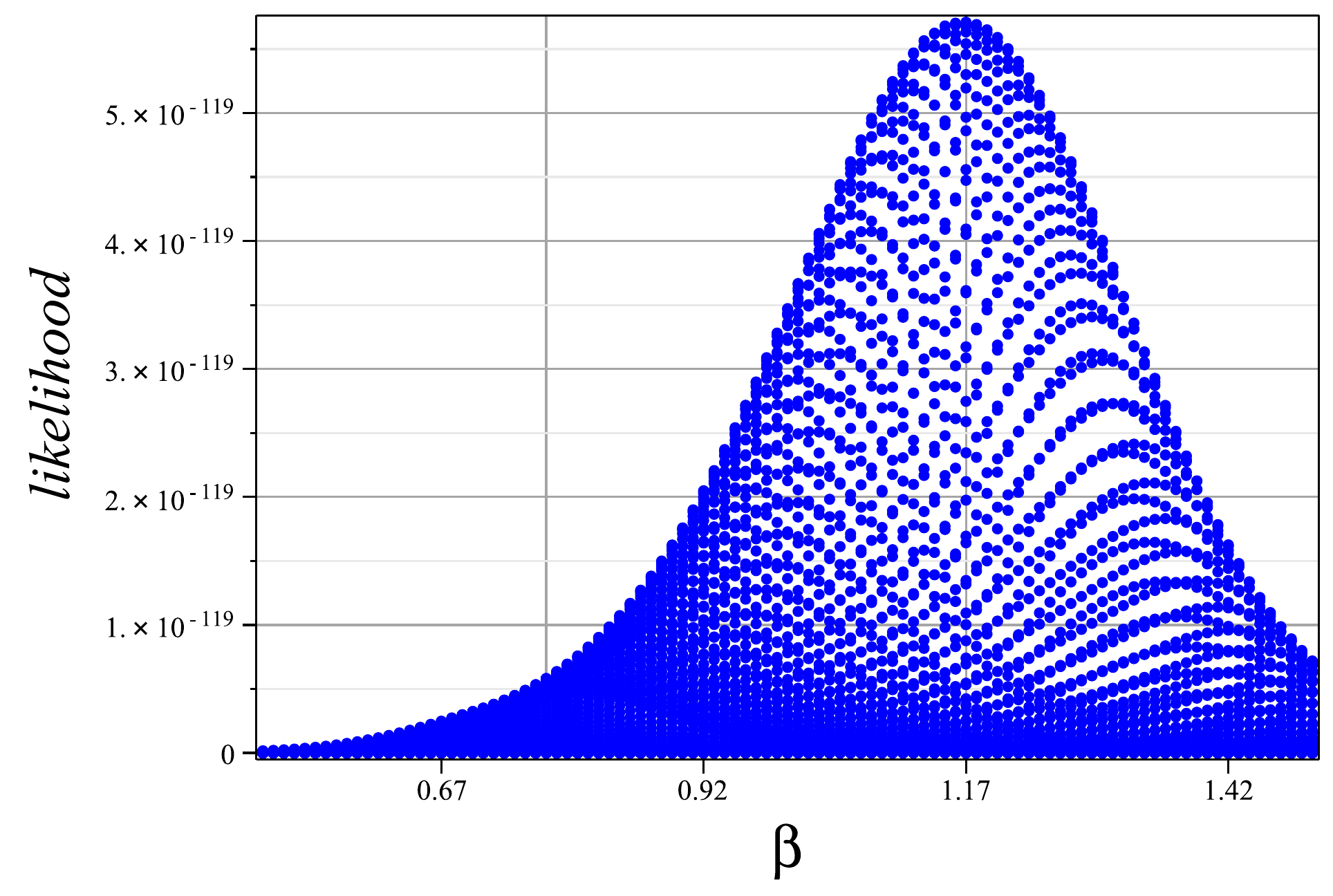}\\
Fig. 1:  The graph of the  one dimensional likelihood distribution for \\parameters $\alpha$ and $\beta$ in exponential case
\end{tabular*}\\
\begin{tabular*}{2.5 cm}{cc}
\includegraphics[scale=.4]{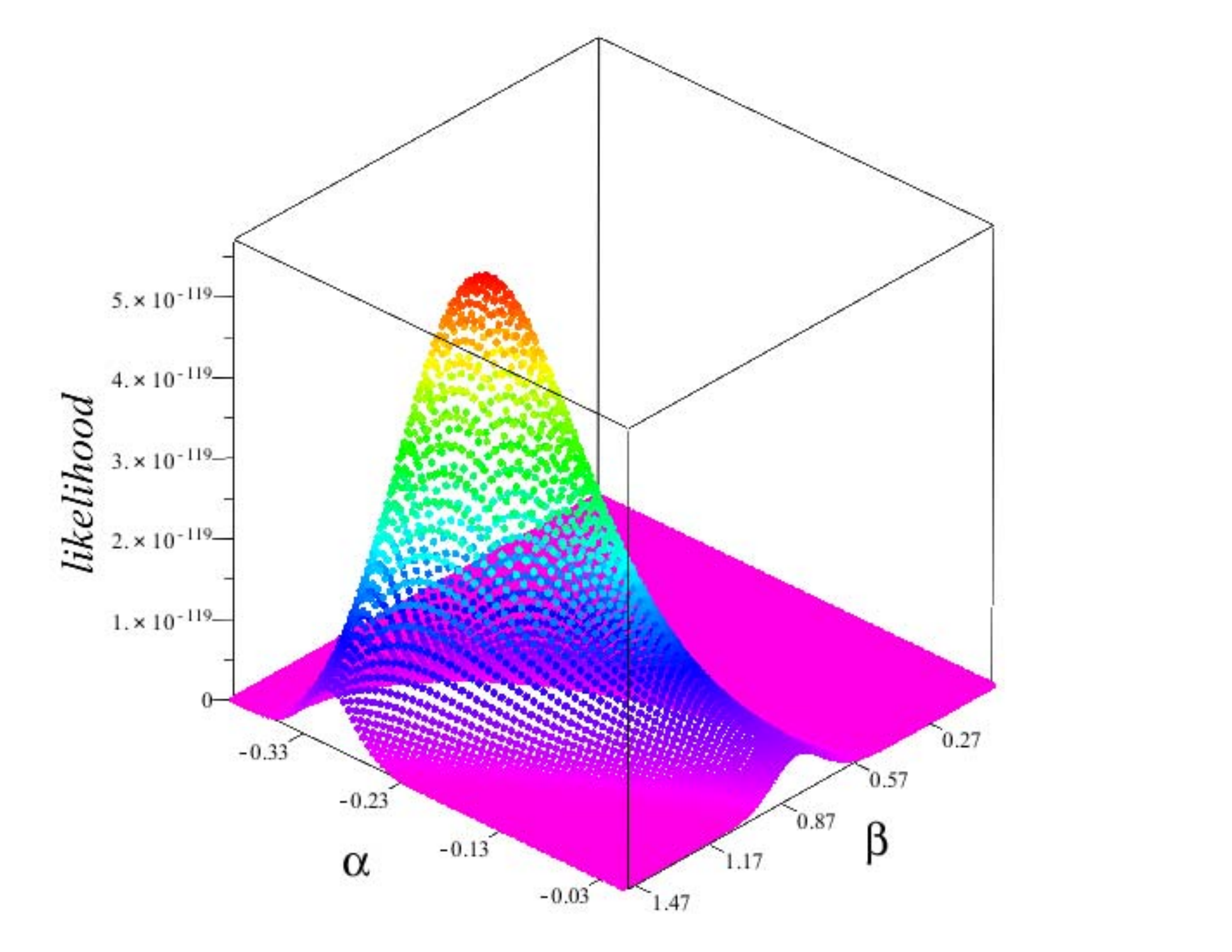}\hspace{0.1 cm}\includegraphics[scale=.35]{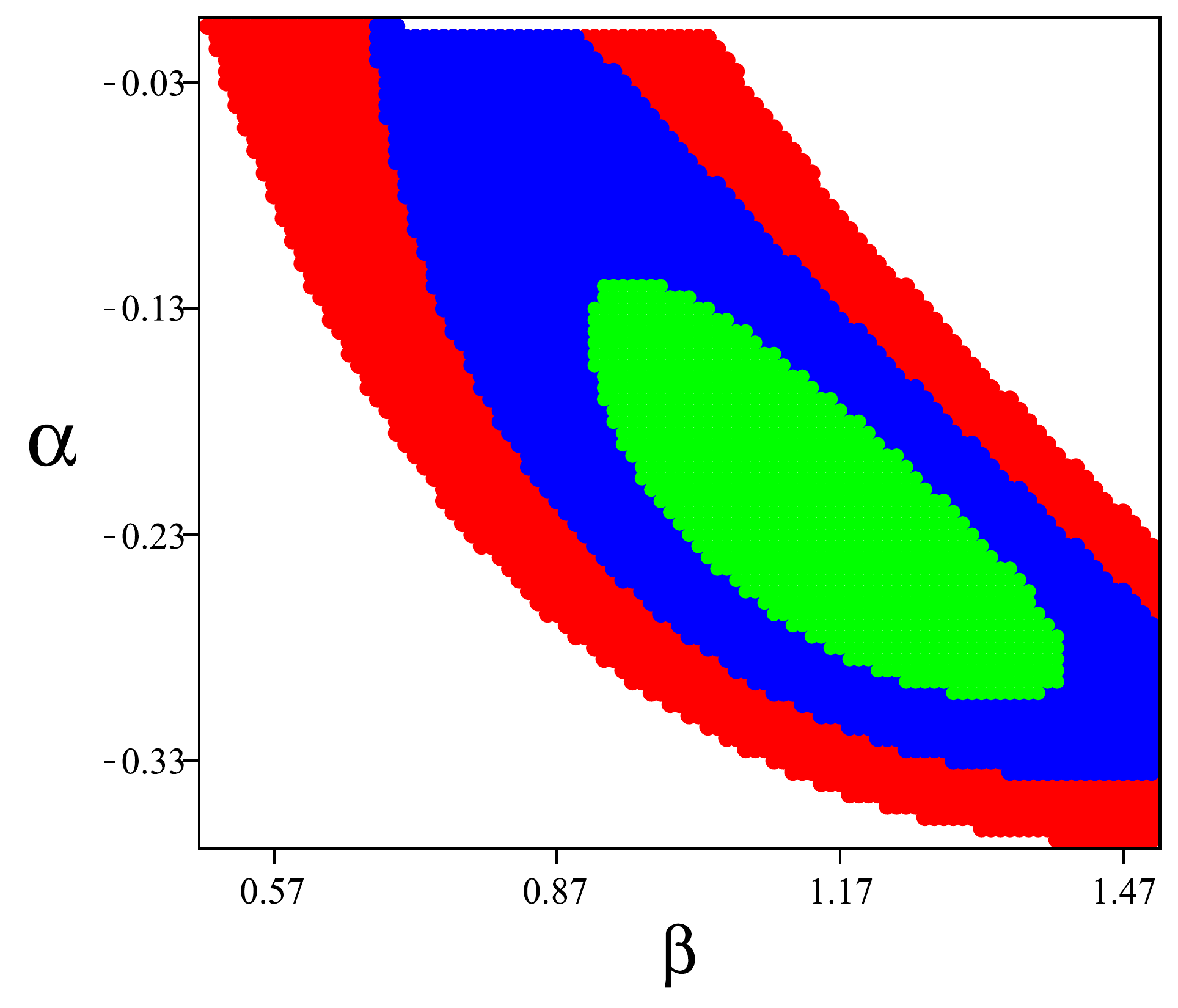}\\
Fig. 2:  The graph of the  two dimensional likelihood distribution and confidence level for \\ parameters $\alpha$ and $\beta$ in exponential case
\end{tabular*}\\
\begin{tabular*}{2.5 cm}{cc}
\includegraphics[scale=.35]{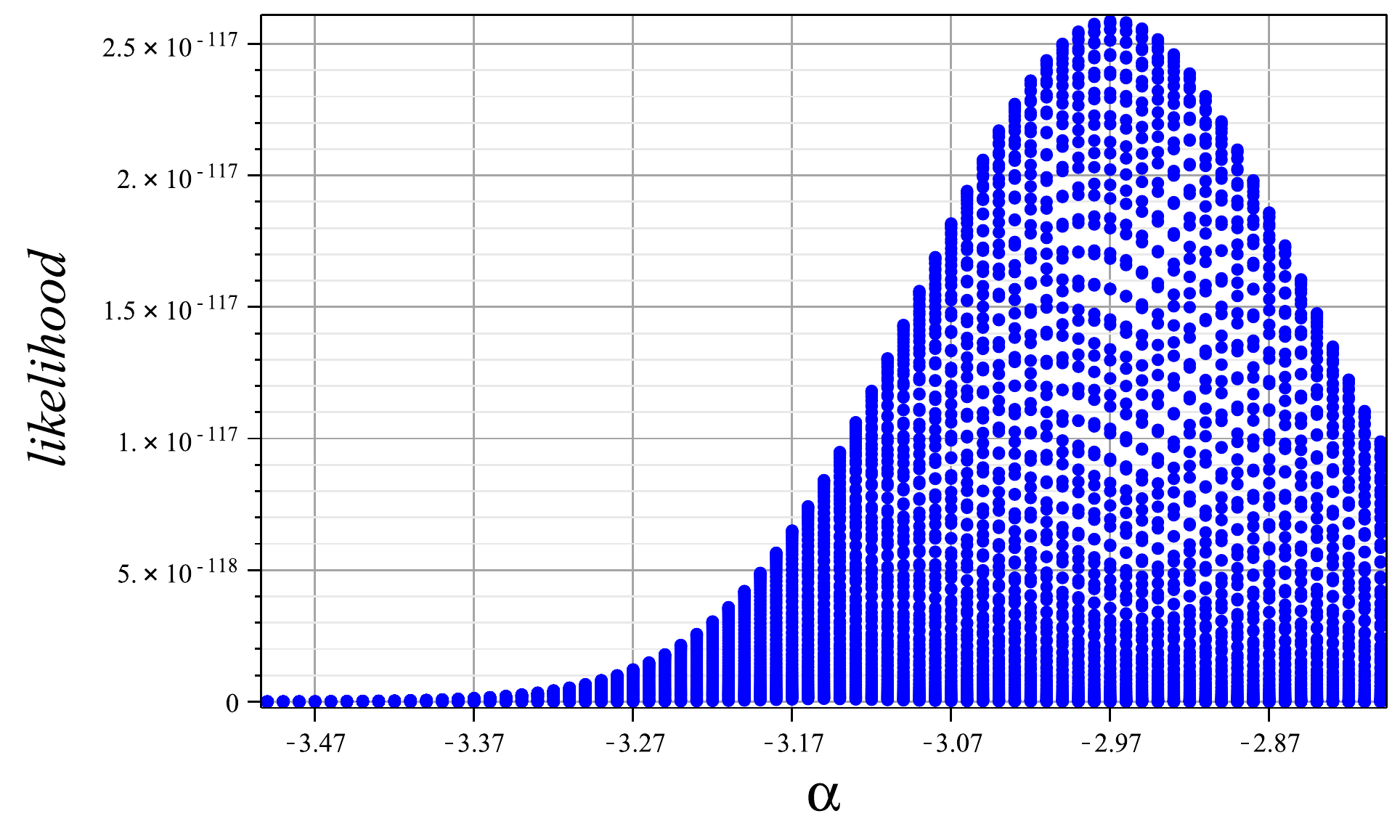}\hspace{0.1 cm}\includegraphics[scale=.35]{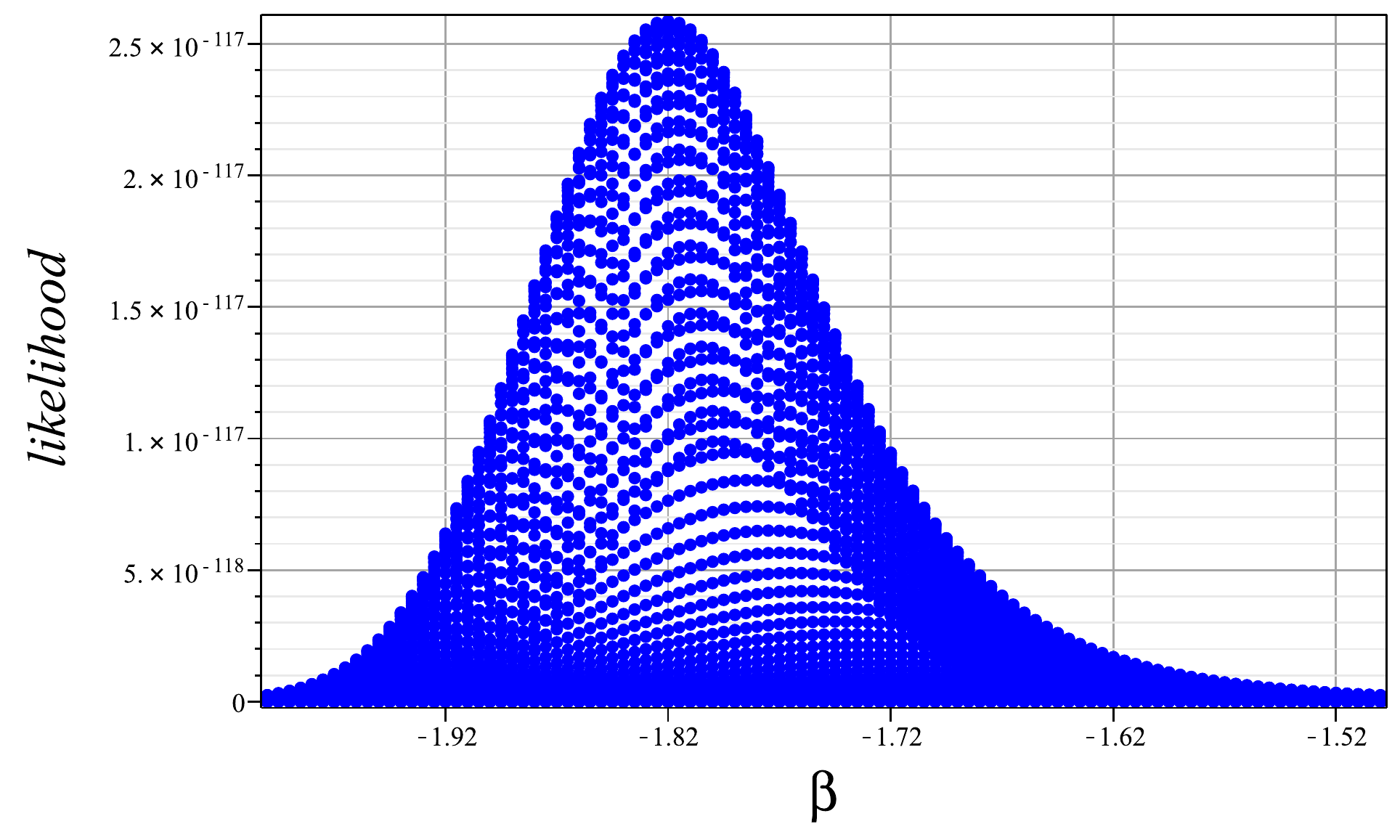}\\
Fig. 3:  The graph of the  one dimensional likelihood distribution for \\ parameters $\alpha$ and $\beta$ in power law case
\end{tabular*}\\
\begin{tabular*}{2.5 cm}{cc}
\includegraphics[scale=.4]{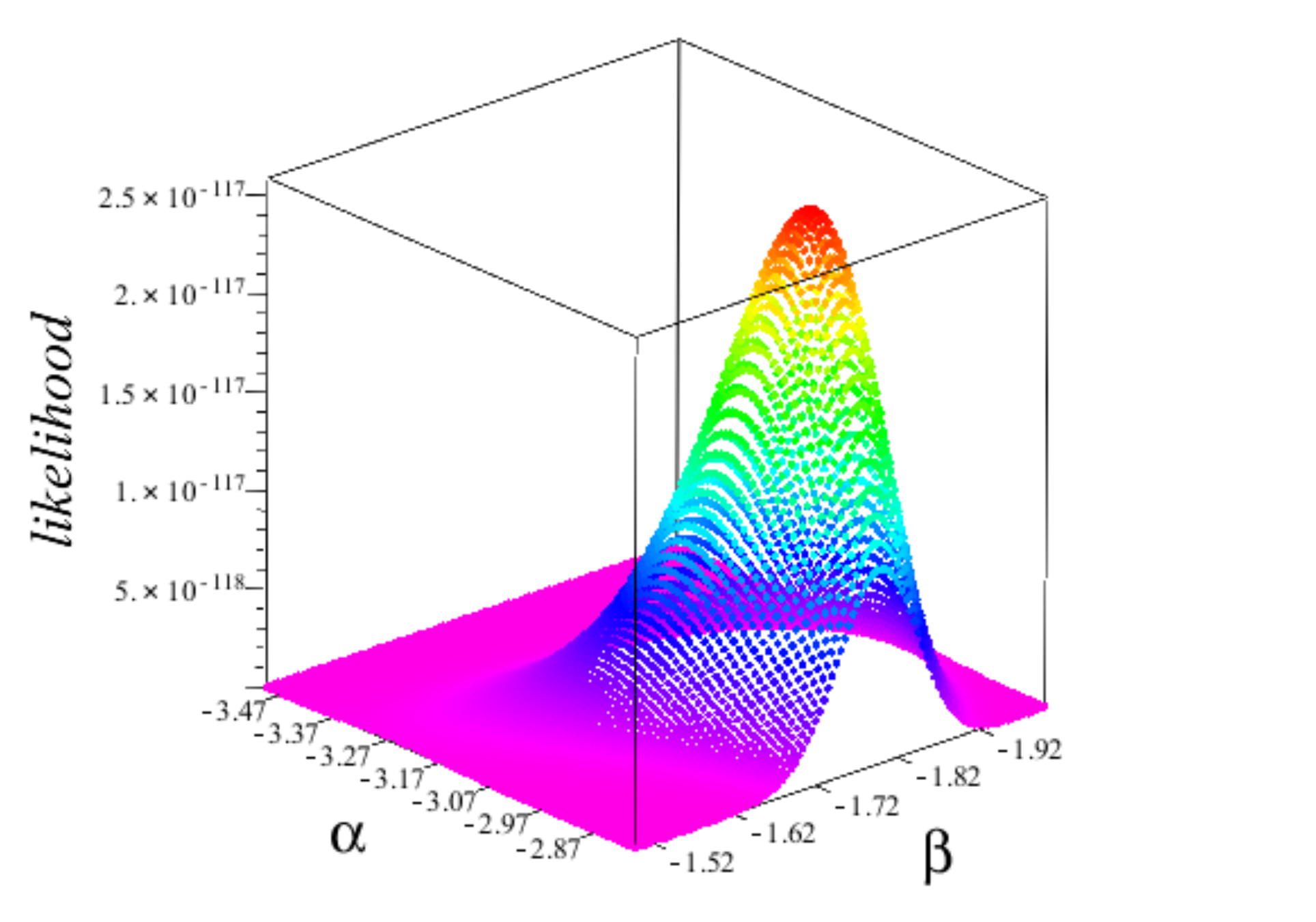}\hspace{0.1 cm}\includegraphics[scale=.35]{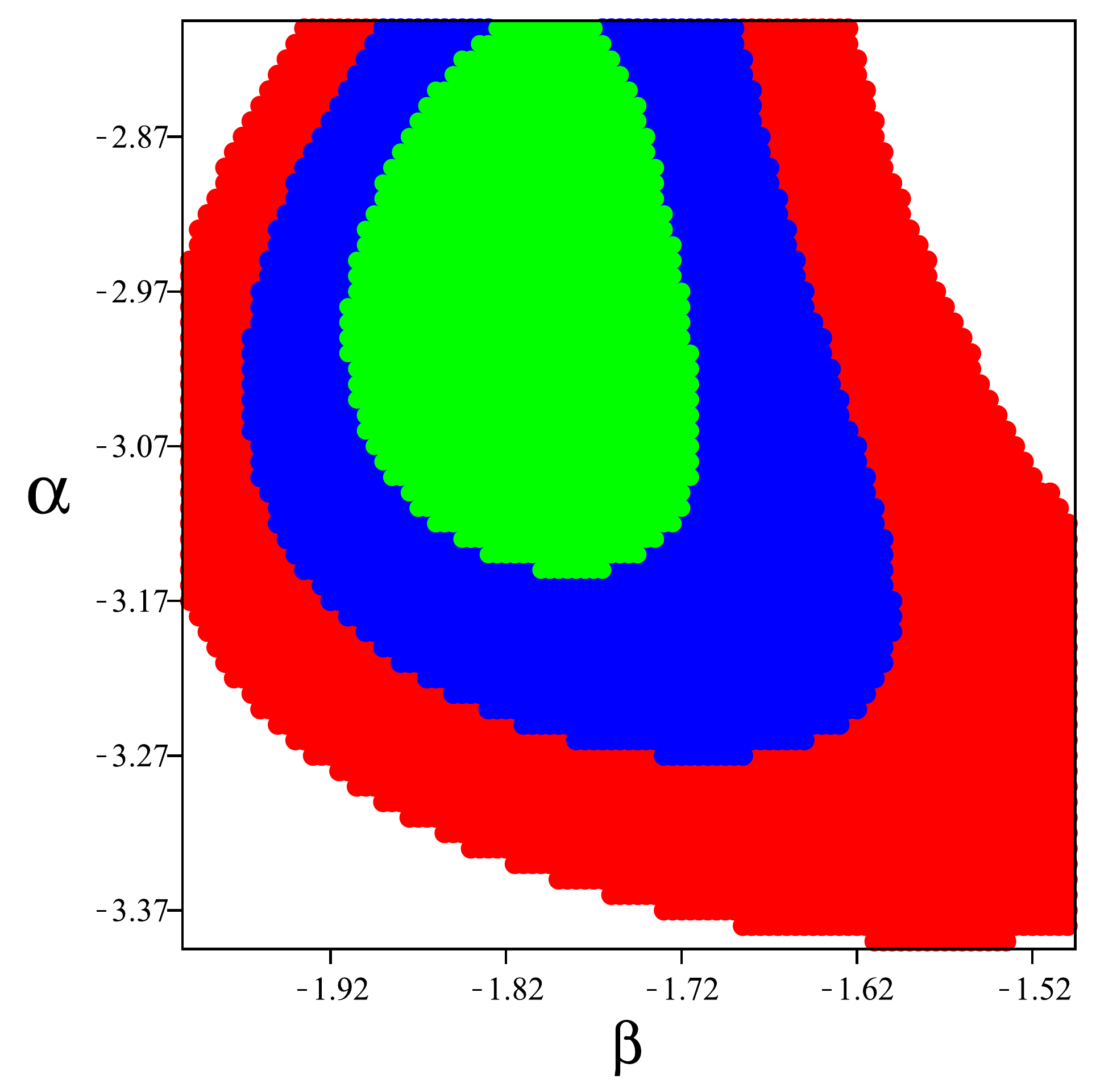}\\
Fig. 4:  The graph of the  two dimensional likelihood distribution and confidence level for \\ parameters $\alpha$ and $\beta$ in power law case
\end{tabular*}\\

The distance modulus, $\mu(z)$, plotted in Fig. 5, in both cases, are best fitted with the most recent SNe Ia observational data for the model parameters and initial conditions using $\chi^2$ method.

\begin{tabular*}{2.5 cm}{cc}
\includegraphics[scale=.4]{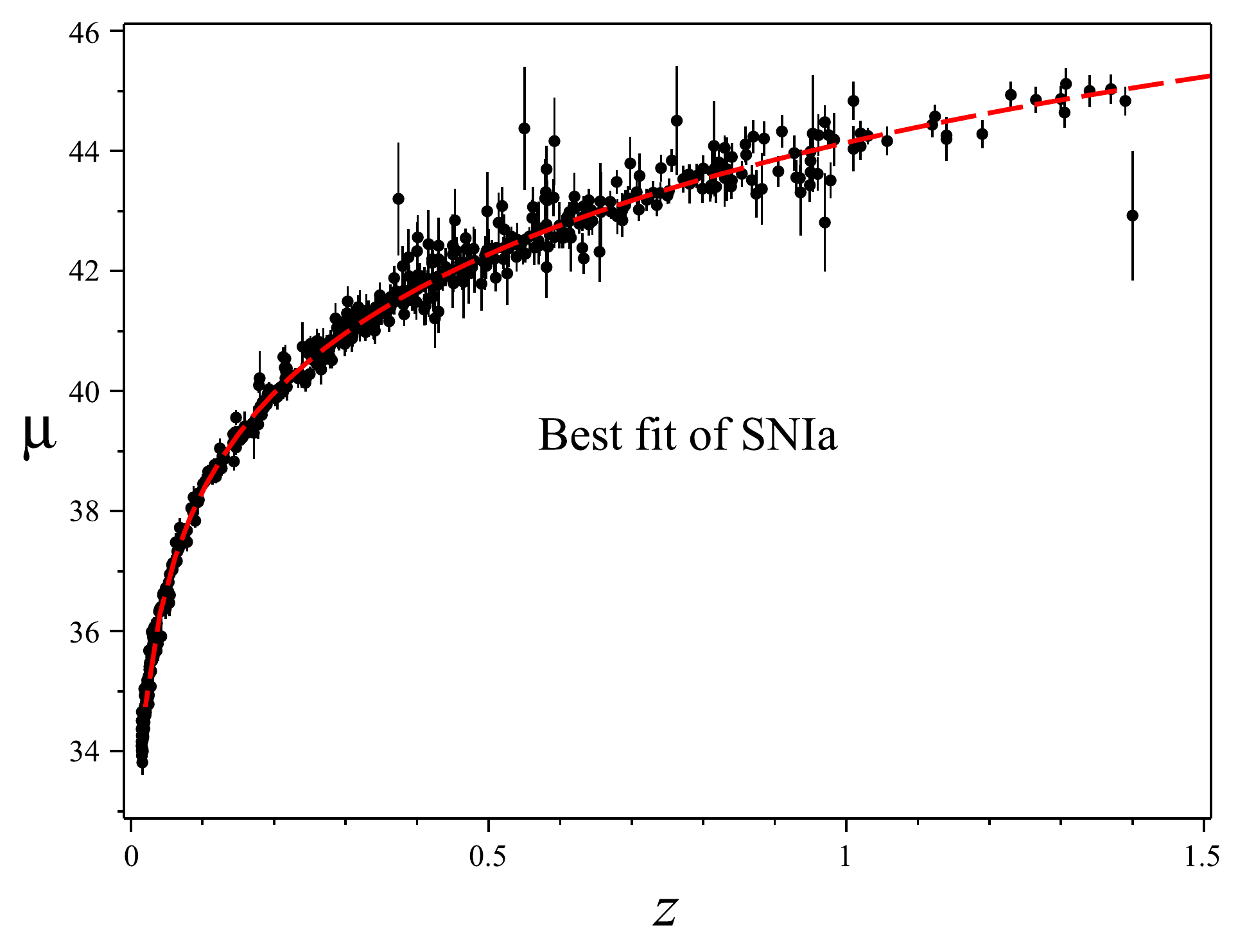}\hspace{0.1 cm}\includegraphics[scale=.4]{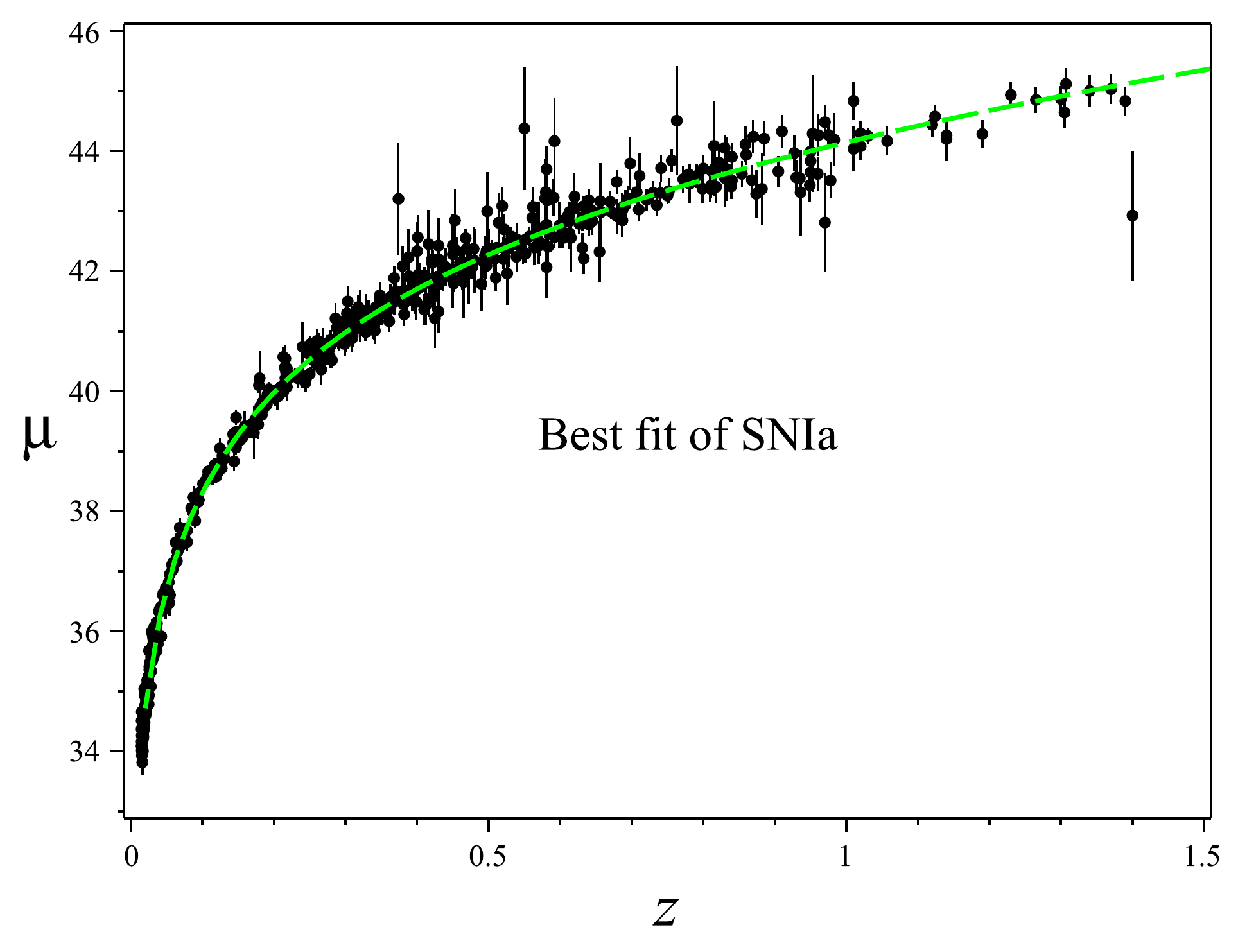}\hspace{0.1 cm}\\
Fig. 5: The best-fitted distance modulus $\mu(z)$ plotted as function of redshift for $F(\phi)$ and $V(\phi)$ \\left) in exponential form,  right) in power law form
\end{tabular*}\\

\section{Cosmological parameters}

In order to understand the behavior of the universe and its dynamics we need to study the cosmological parameters. We have best fitted our model with the current observational data by the distance modulus test. The cosmological parameters analytically and/or numerically have been investigated by many authors for variety of cosmological models. Simultaneously best fitting the model with the observational data gives us a better understanding of the solutions and the dynamics of these parameters. Among cosmological parameters, the effective EoS parameter and statefinders are given by
$ w_{eff}=-1-2\dot{H}/3H^{2}$, $r=\ddot{H}/H^{3}-3q-2$ and $s=(r-1)/3(q-\frac{1}{2})$ discussed here, where $q$ in $r$ and $s$ is the deceleration parameter and $\frac{\ddot{H}}{H^{2}}$ in $r$ in terms of new dynamical variables for exponential and power law cases can be obtained by taking derivative of $\dot{H}$.

In Fig. 6, the effective EoS parameters are shown in both cases of exponential functions (left panel) and power law functions (right panel). In both cases the effective EoS parameters are best fitted for the model parameters and initial conditions with the observational data. From the graph, for the exponential case, the EoS parameter approaches zero in high redshift as expected for matter dominated universe. It also shows that at about $z \leq 0.5$ in the past where $\omega_{eff}\leq -1/3$ the universe begins to accelerate which is consistent with observational evidence. The result does not show phantom crossing in the past and present. In the case of power law functions the result is very disappointing. The effective EoS parameter become tangent to the phantom divide line in the past which is not observationally justified. Note that in both cases the trajectories are best fitted for the model parameters and initial conditions with the observational data. The satisfactory in the exponential case and discrepancy in power law case demonstrate an advantage for the exponential behavior of these functions. \\

\begin{tabular*}{2.5 cm}{cc}
\includegraphics[scale=.35]{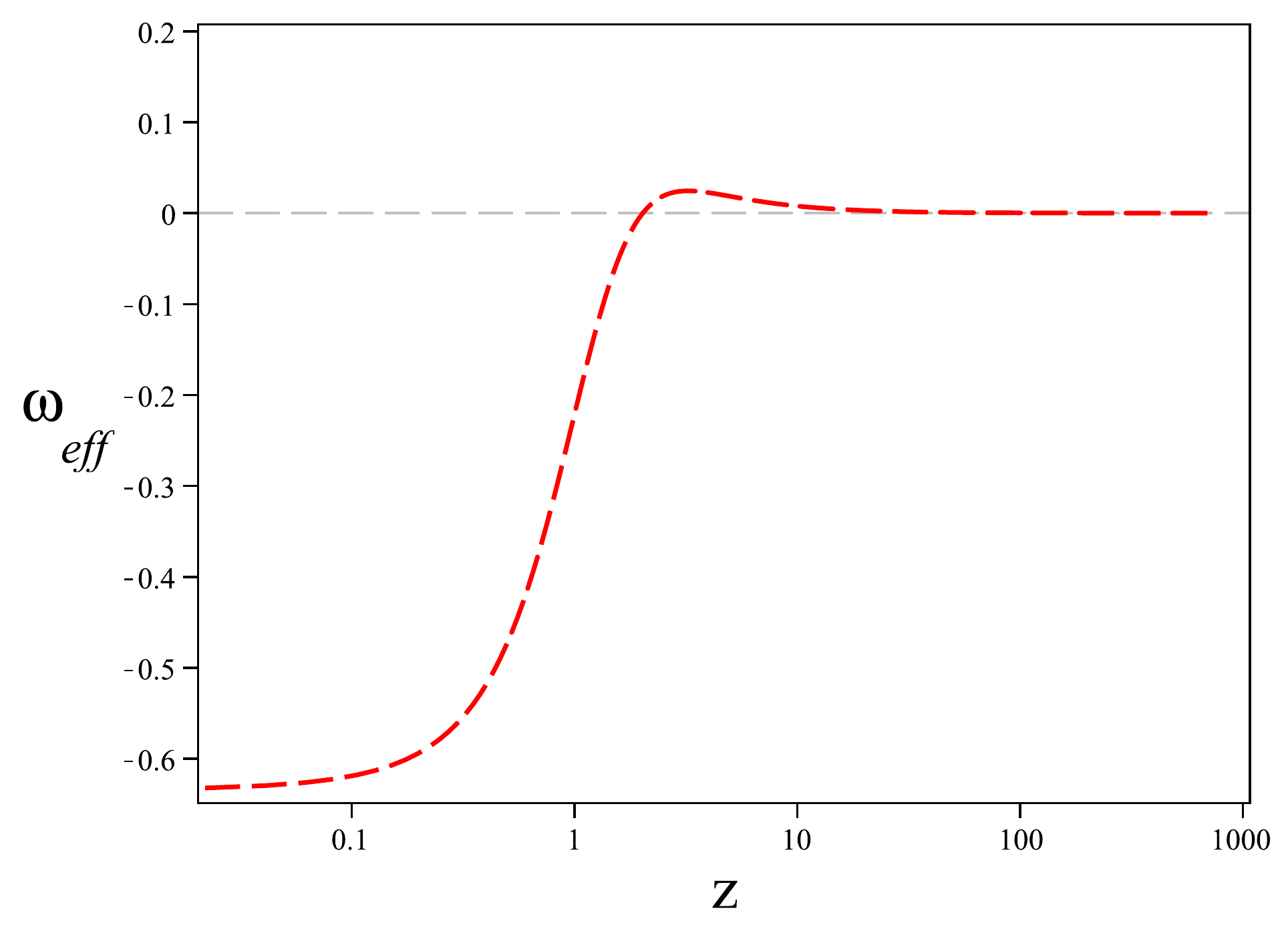}\hspace{0.1 cm}\includegraphics[scale=.35]{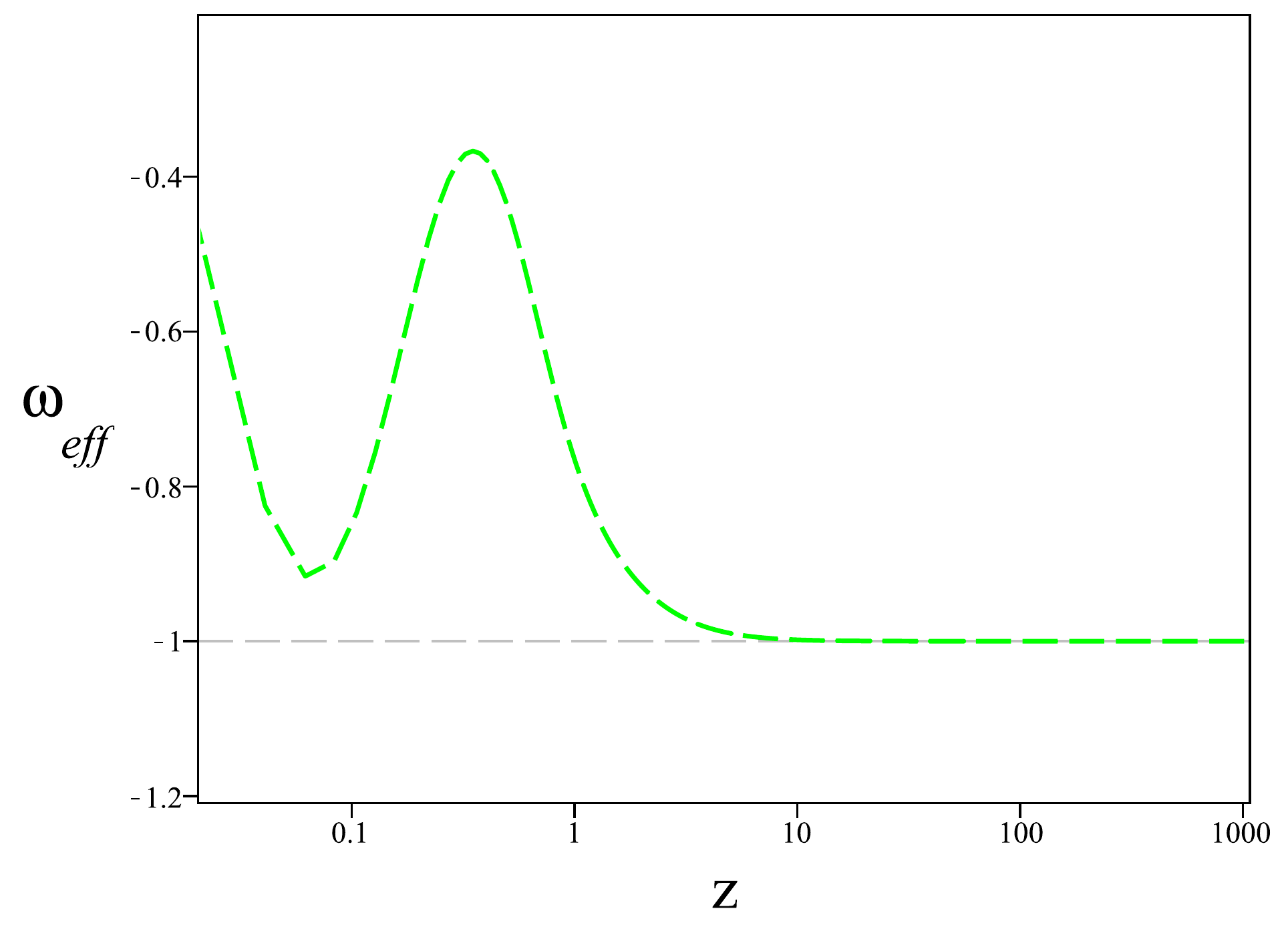}\hspace{0.1 cm}\\
Fig. 6: The best-fitted effective equation of state parameter, $\omega_{eff}$, plotted as function of \\redshift for $F(\phi)$ and $V(\phi)$ left) in exponential form,  right) in power law form
\end{tabular*}\\

Fig 7 shows the best-fitted trajectories of the statefinder diagrams $\{r,s\}$ in both exponential and power law cases. From the graph it can be seen that both best-fitted trajectories passed LCDM state with $\{r, s\}=\{1, 0\}$ sometimes in the past.
The current value of the best fitted trajectory in exponential case and its location with respect to
the LCDM state can also be observed in the $\{r,s\}$ diagram. However, in power law case, since the current location of the trajectory is far from LCDM state, in order to find its location and value in Fig 8 we depict the corresponding dynamical behavior of the satefinder  $\{r,s\}$ against $N=-ln(1+z)$ in both cases of exponential and power law. From Fig 7, we see that the current value of the statefinder $\{r, s\}$ in exponential case is $\{0.48, 0.19\}$, close to the LCDM state. From Fig 8 in power law case it has been shown that the current value of trajectory is $\{-42, -25\}$, relatively far from LCDM in statefinder diagram. This analysis shows another advantage of considering exponential behavior for the functions $F(\phi)$ and $V(\phi)$ over power law behavior. This is because the recent observational data confirms LCDM as the standard cosmological model in the current epoch.

\begin{tabular*}{2.5 cm}{cc}
\includegraphics[scale=.35]{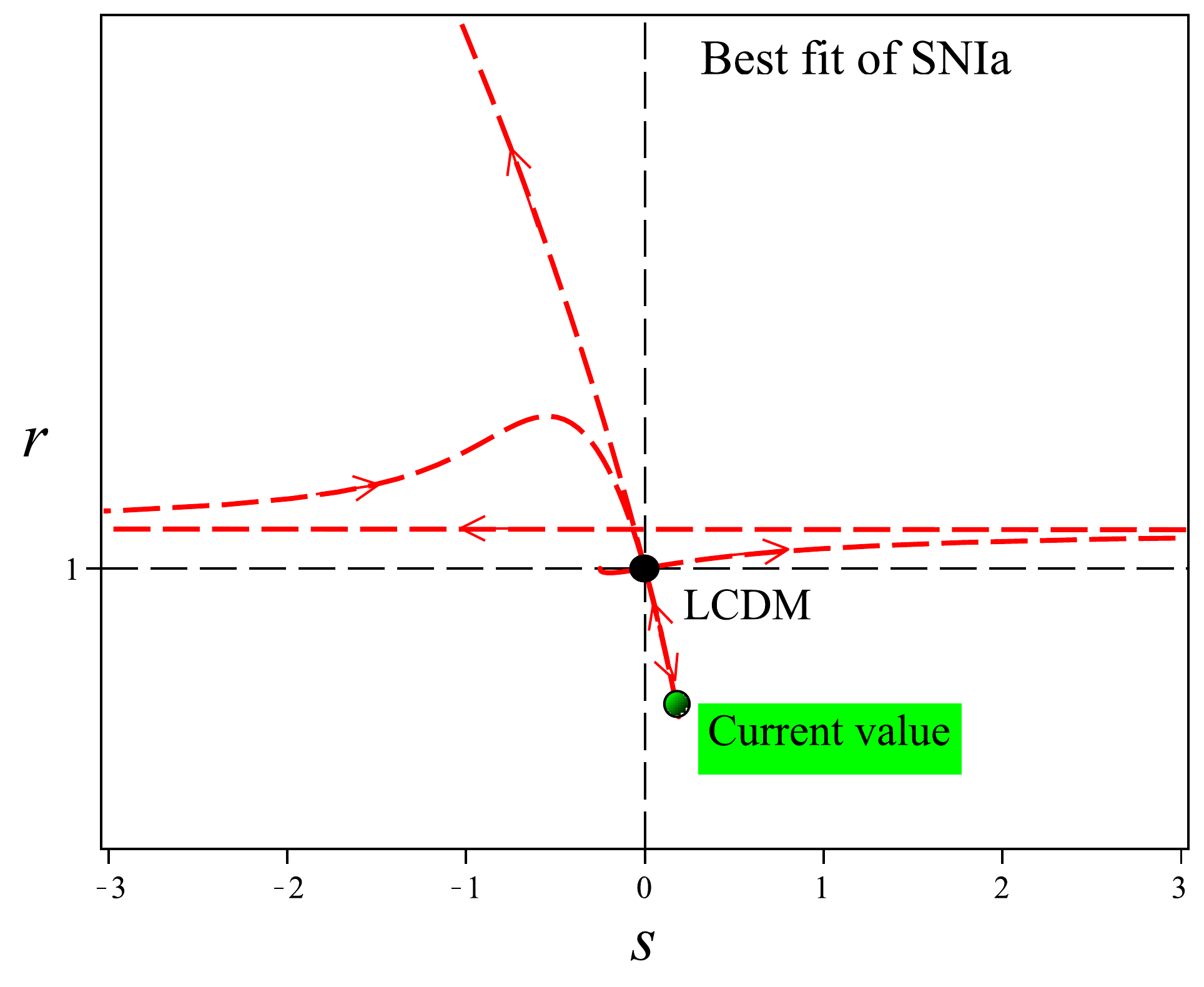}\hspace{0.1 cm}\includegraphics[scale=.35]{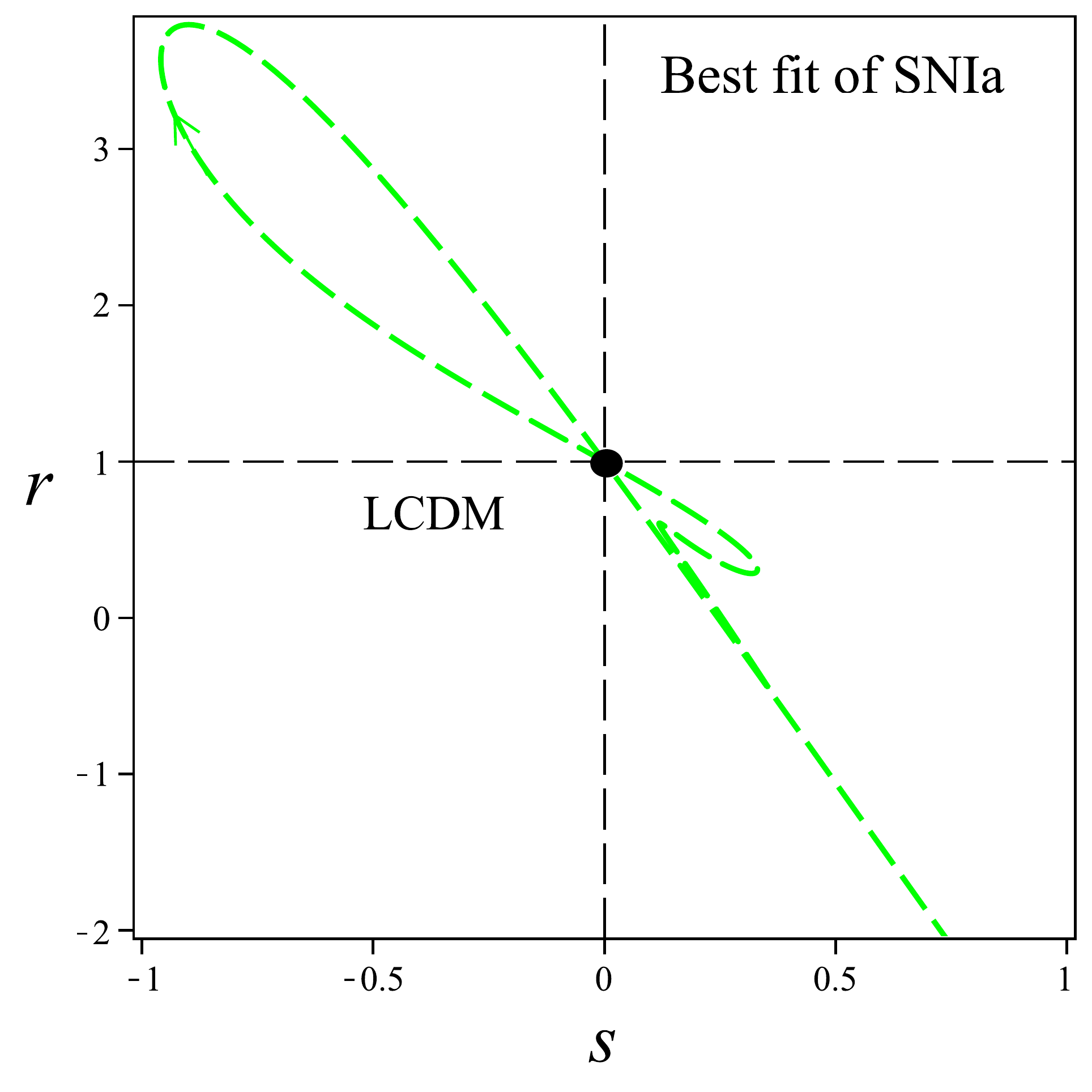}\hspace{0.1 cm}\\
Fig. 7: The best-fitted statefinder parameter $\{r, s\}$ for $F(\phi)$ and $V(\phi)$ \\left) in exponential form,  right) in power law form
\end{tabular*}\\

\begin{tabular*}{2.5 cm}{cc}
\includegraphics[scale=.35]{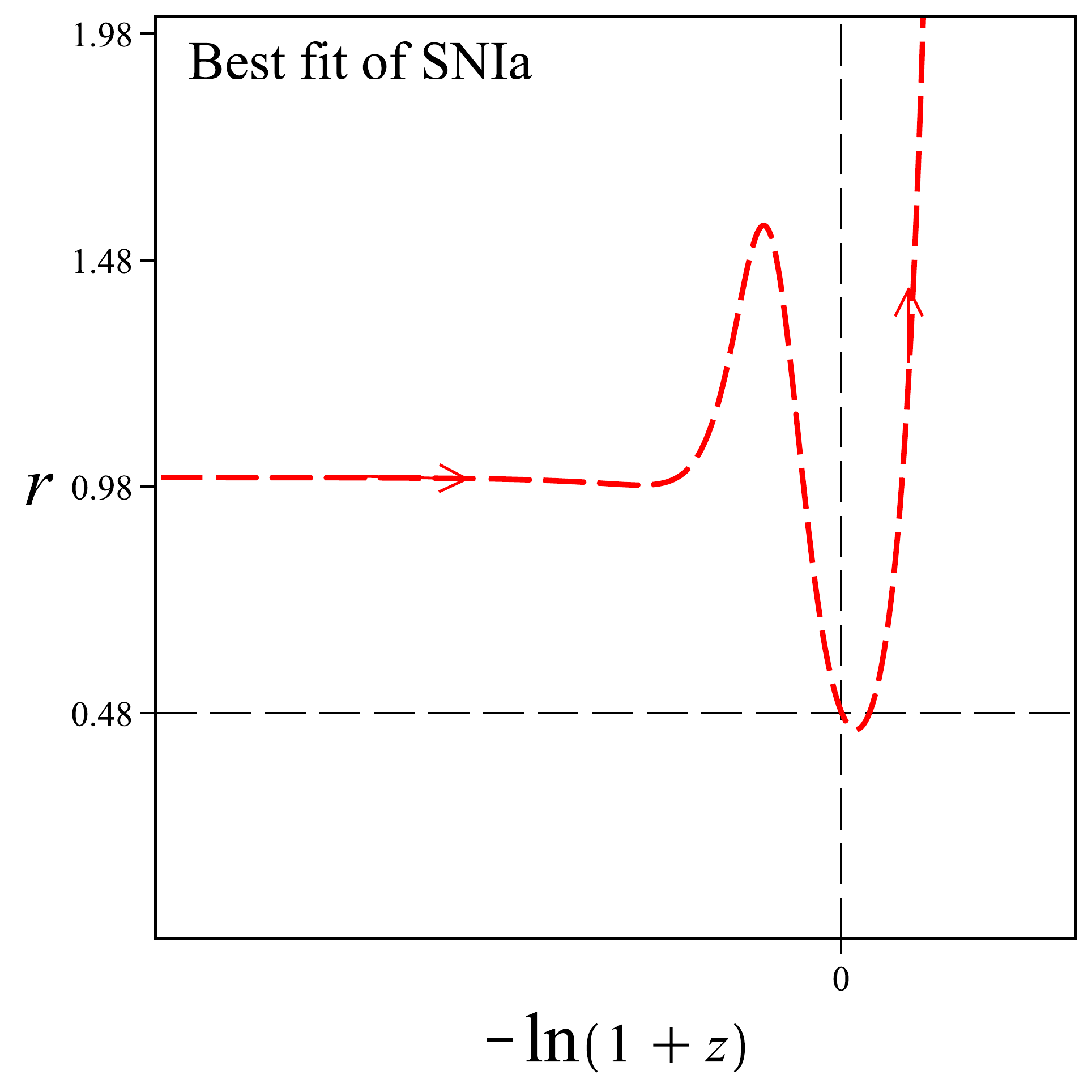}\hspace{0.1 cm}\includegraphics[scale=.35]{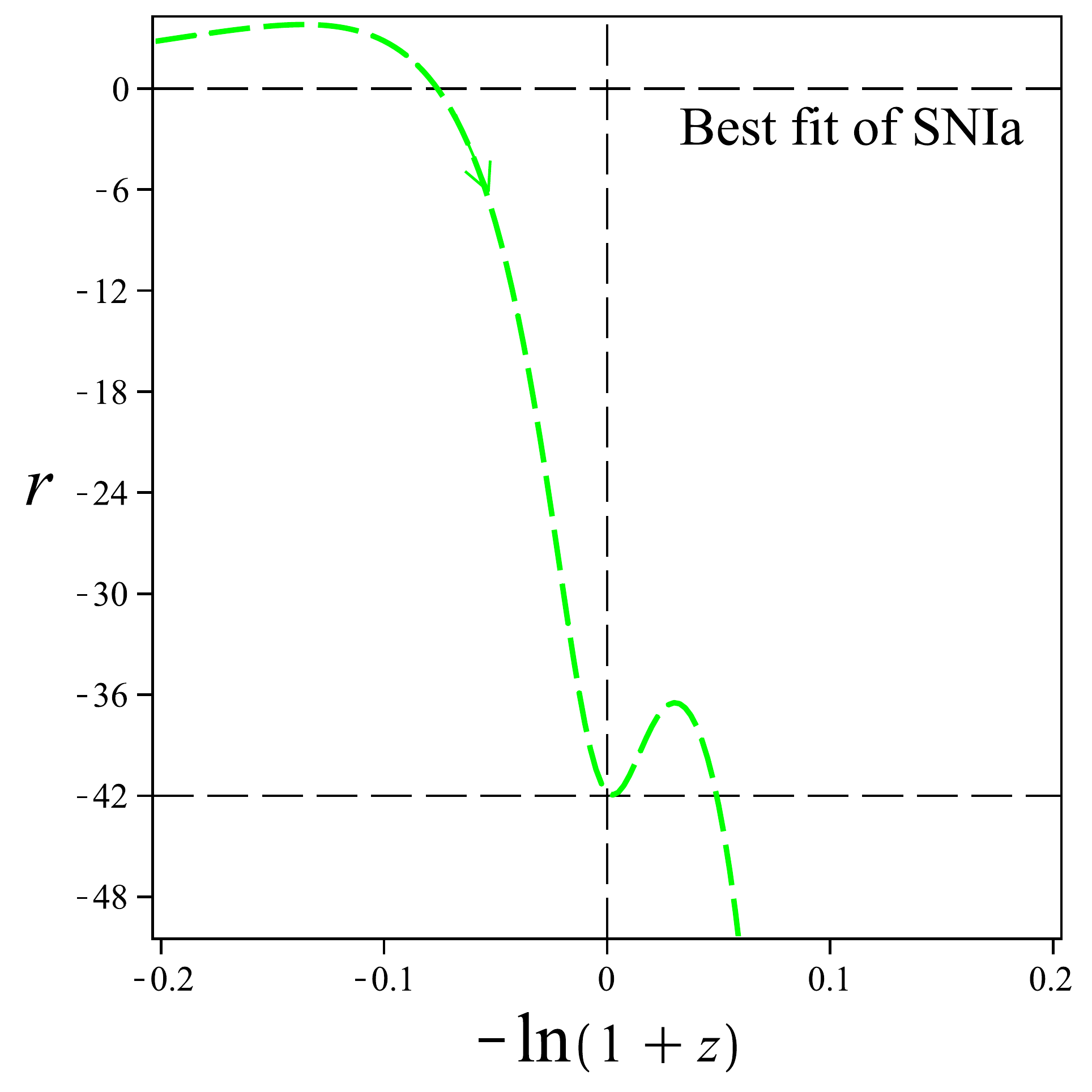}\hspace{0.1 cm}\\
\includegraphics[scale=.42]{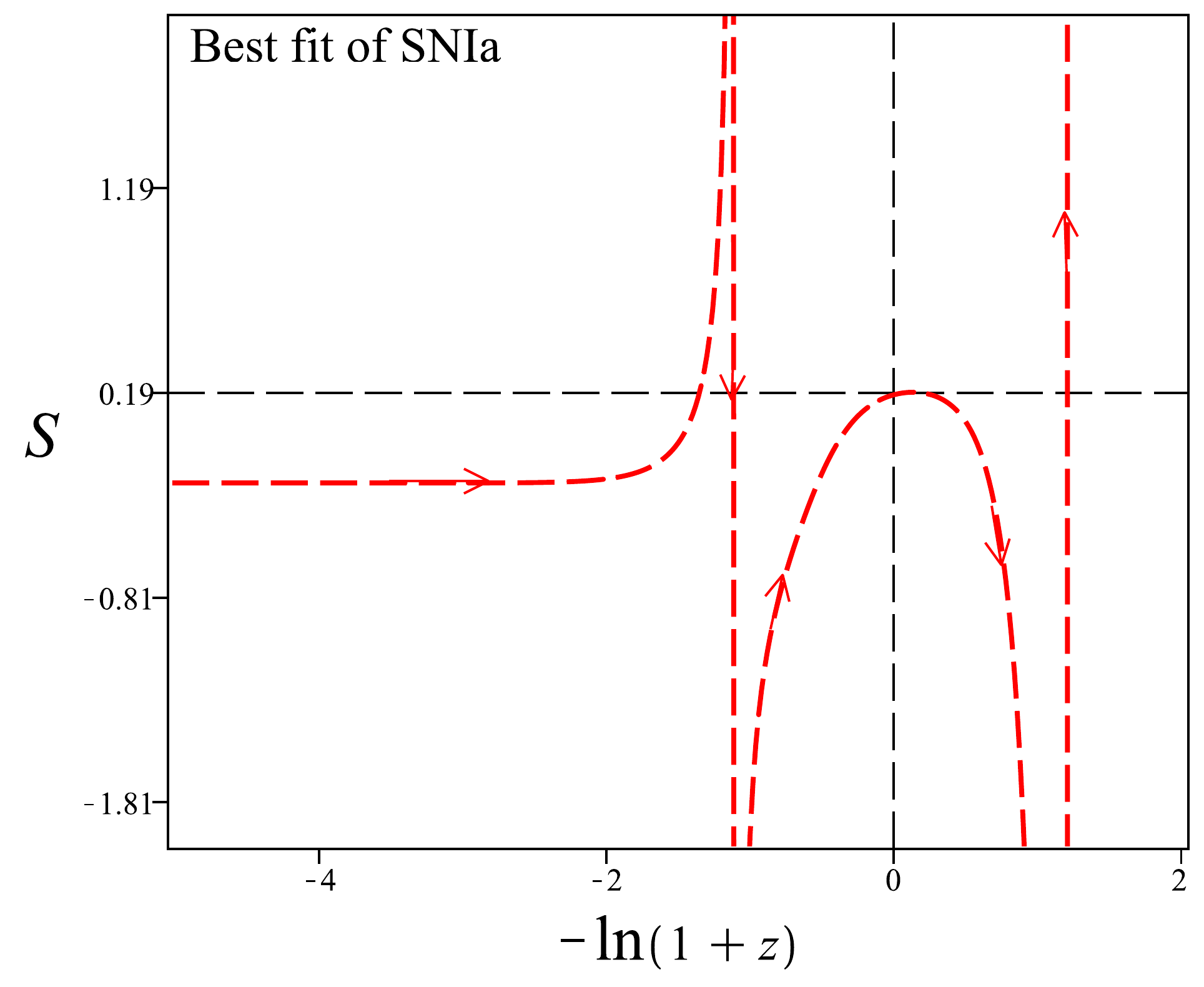}\hspace{0.1 cm}\includegraphics[scale=.35]{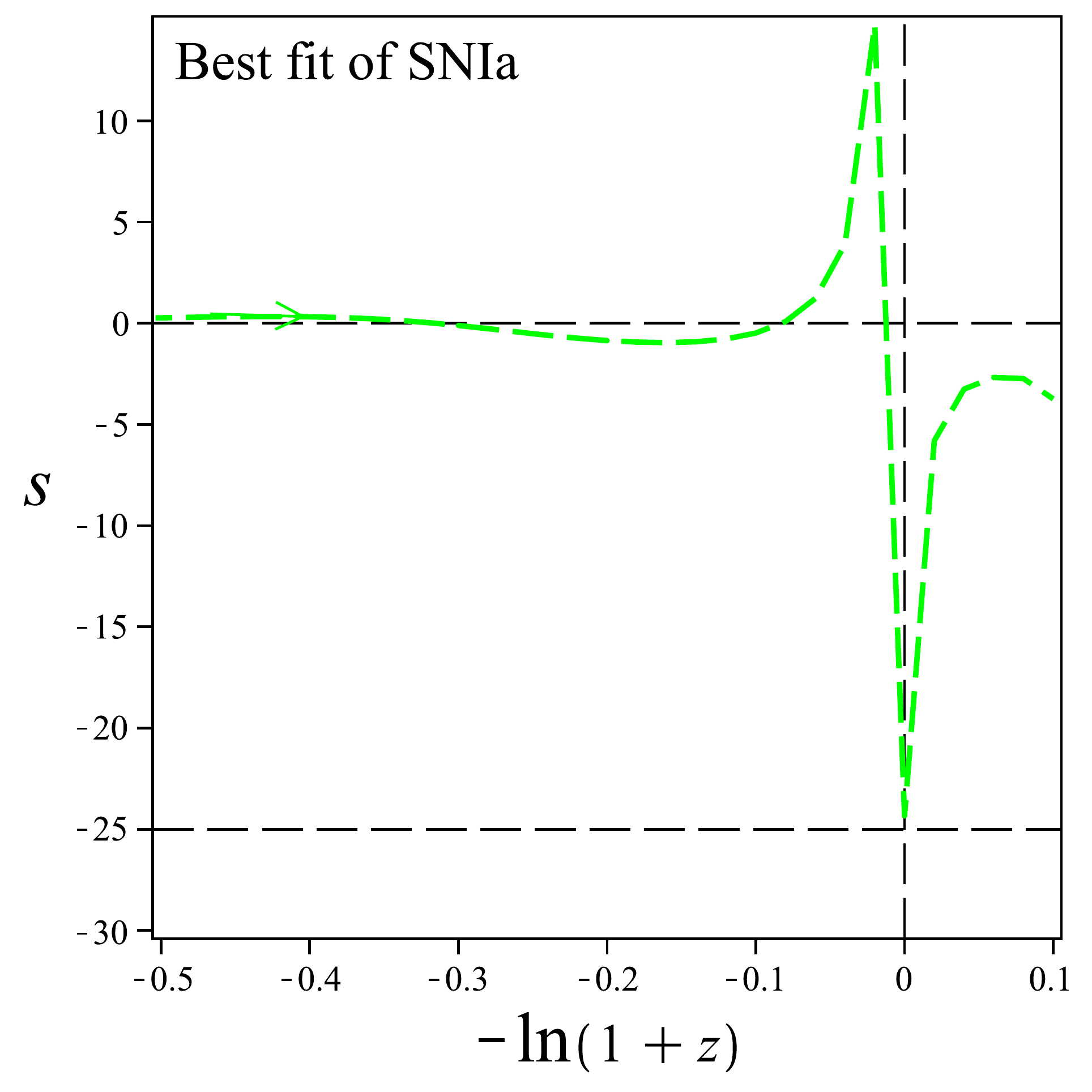}\hspace{0.1 cm}\\
Fig. 8: The dynamics of the statefinder parameters $r$ and $s$ with respect to $N=-ln (1+z)$ for \\ $F(\phi)$ and $V(\phi)$ left) in exponential form,  right) in power law form
\end{tabular*}\\

We also reconstructed the best-fitted function $\dot{\phi}$ in both exponential and power law cases. From Fig 9, we see that the $\dot{\phi}$ trajectory for the best fitted model parameters in exponential case shows
monotonic increasing behavior in redshift, whereas in power law case shows a constant behavior. Both cases have similar behavior in the near past up to the redshift $z\simeq 0.9$.

\begin{tabular*}{2.5 cm}{cc}
\includegraphics[scale=.4]{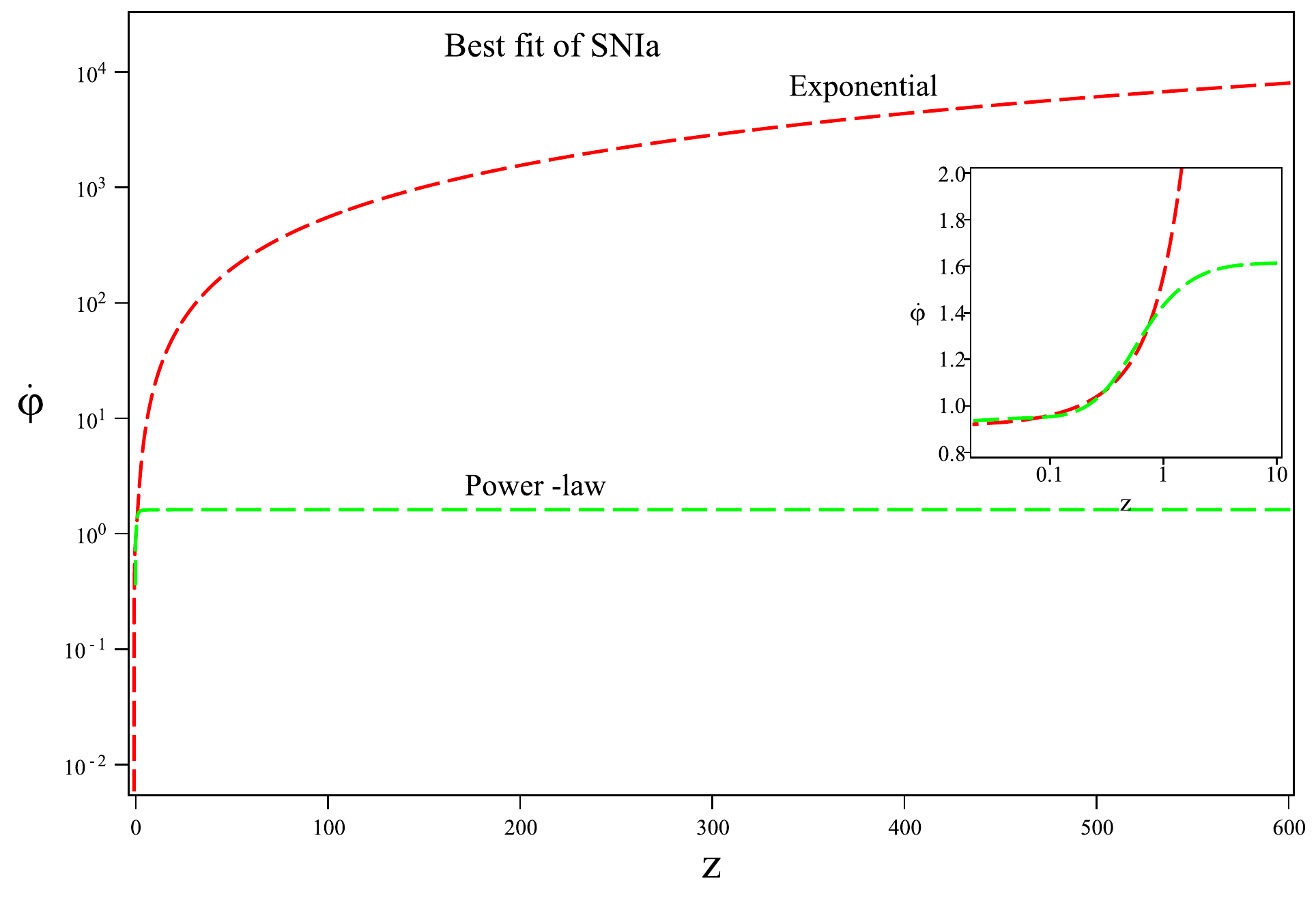}\hspace{0.1 cm}\\
Fig. 9: The best-fitted reconstructed $\dot{\phi} $plotted as function of redshift for $F(\phi)$ and $V(\phi)$ \\ left) in exponential form,  right) in power law form
\end{tabular*}\\

\section{Summary and remarks}

This paper is designed to study the dynamics of scalar tensor cosmology with tachyonic potential and non minimally coupled scalar field function with the curvature and matter lagrangian. The model characterized by the scalar field functions $F(\phi)$ and $V(\phi)$. We investigate two forms of the exponential and power law behavior for these functions. To solve the field equations, in a new approach, we simultaneously best-fit the model parameters and initial conditions with the observational data using $\chi^2$ method. The advantage of this approach is that the obtained solutions are observationally verified and thus physically more promising.

We then study the cosmological parameters such as effective EoS and statefinder parameters
for the model in terms of the  best-fitted model parameters and initial conditions. The result shows that the best-fitted effective EoS parameter in case of exponential function, while does not cross the cosmological divide line in the near and far past, exhibits an observationally verified behavior in the past ($\omega_{eff}=0$, matter dominated universe) . In addition its current value is within the range of observationally accepted values. On the contrary, in power law case, the dynamics of the best-fitted parameter is very unsatisfactory as the result shows an accelerating universe at higher redshifts. The best fitted statefinder parameters show that in exponential case the current state of the universe is very close to LCDM, whereas in the power law case it is relativity far from LCDM in the statefinder diagram. As noted, this behavior shows another advantage of exponential behavior of the functions over power law one. Finally, the best fitted scalar field function, $\dot{\phi}$, in exponential case displays an interesting behavior, smoothly decreasing function

\end{document}